# Reasoning-Driven Design of Single Atom Catalysts via a Multi-Agent Large Language Model Framework

*Dong Hyeon Mok[1], Seoin Back[2,3,4,*], Victor Fung[5,*] and Guoxiang Hu[6,7,*]*

[1]Department of Chemical and Biomolecular Engineering, Institute of Emergent Materials, Sogang University, Seoul 04107, Republic of Korea

[2]KU-KIST Graduate School of Converging Science and Technology, Korea University, Seoul 02841, Republic of Korea

[3]Department of Integrative Energy Engineering, Korea University, Seoul 02841, Republic of Korea

[4]Institute for Multiscale Matter and Systems (IMMS), Ewha Womans University, Seoul 03760, Republic of Korea

[5] School of Computational Science and Engineering, Georgia Institute of Technology, Atlanta, GA 30332, USA.

[6] School of Materials Science and Engineering, Georgia Institute of Technology, Atlanta, GA 30332, USA.

[7] School of Chemistry and Biochemistry, Georgia Institute of Technology, Atlanta, GA 30332, USA.

AUTHOR INFORMATION

**Corresponding Author**

E-mail: sback@korea.ac.kr, victorfung@gatech.edu, emma.hu@mse.gatech.edu

**Abstract**

Large language models (LLMs) are becoming increasingly applied beyond natural language processing, demonstrating strong capabilities in complex scientific tasks that traditionally require human expertise. This progress has extended into materials discovery, where LLMs introduce a new paradigm by leveraging reasoning and in-context learning, capabilities absent from conventional machine learning approaches. Here, we present a Multi-Agent-based Electrocatalyst Search Through Reasoning and Optimization (MAESTRO) framework in which multiple LLMs with specialized roles collaboratively discover high-performance single atom catalysts for the oxygen reduction reaction. Within an autonomous design loop, agents iteratively reason, propose modifications, reflect on results and accumulate design history. Through in-context learning enabled by this iterative process, MAESTRO identified design principles not explicitly encoded in the LLMs' background knowledge and successfully discovered catalysts that break conventional scaling relations between reaction intermediates. These results highlight the potential of multi-agent LLM frameworks as a powerful strategy to generate chemical insight and discover promising catalysts.

# 1. Introduction

Identifying new materials that deliver high catalytic activity, selectivity, and long-term stability has remained a significant and ongoing scientific challenge, with major implications for improving energy efficiency, lowering industrial costs, and enabling more sustainable chemical processes. Over time, materials discovery strategies have evolved beyond purely experimental approaches and first-principles simulations, such as density functional theory (DFT) calculations[1-3], with machine learning (ML) becoming one of the dominant paradigms in the field[4-9]. Early ML-based approaches primarily relied on high-throughput screening[10-13], in which large datasets were generated in advance and property prediction models were used to filter promising candidates, thereby reducing the computational and experimental cost of subsequent validation. However, as the chemical search space expanded and performance requirements became increasingly demanding, such screening-based strategies encountered limitations related to data availability, computational cost, and scalability[14, 15]. To address these challenges, inverse design strategies were introduced, in which materials are designed starting from target properties rather than discovered through forward searches[16-19]. Approaches based on global optimization[20, 21] and generative models[22-24] have demonstrated notable success in the efficient discovery of promising materials. Despite these advances, data-driven methods remain inherently constrained by their training dataset[25, 26]. While they can discover novel materials within established structure-property relationships, they often lack explanatory power regarding why a particular material is selected and struggle to identify materials governed by unknown physical mechanisms outside the learned domain. As a result, the discovery of fundamentally new, physics-driven design principles has largely remained dependent on human intuition and intervention.

Recent advances in large language models (LLMs) offer a fundamentally different avenue for inverse design. Originally developed for natural language processing, LLMs trained on massive text corpora have demonstrated a capacity for human-like reasoning[27-29]. In chemistry and materials science, LLMs have begun to move beyond simple data mining toward roles involving experimental planning[30], synthesis guidance[31], simulation tool orchestration[32] and chemically informed decision-making tasks that were previously exclusive to human experts[33-36]. Crucially, LLMs possess the capability for in-context learning, whereby they adapt their reasoning based on prior interactions and accumulated context without explicit parameter updates[37]. This enables LLMs to acquire task-specific principles and uncover new insights not explicitly encoded in their background knowledge, particularly when embedded within iterative decision-making loops[38]. Furthermore, recent research has shifted from assigning multiple roles to a single LLM toward multi-agent frameworks, whereby several LLM-based agents, each specialized for a specific task, interact and collaborate[39, 40]. Such systems provide a natural foundation for complex scientific workflows that require hypothesis formulation, reflection and memory.

In this work, we investigate whether an LLM-based multi-agent framework can be successfully applied to heterogeneous electrocatalysis, a relatively narrow yet intrinsically complex domain. While LLMs have recently been employed in catalyst-related studies for task such as data mining[41], synthesis planning[42], property prediction[43] and catalyst generation[44], approaches that explicitly leverage LLM reasoning to design high-performance catalysts remain scarce compared to other systems. To address this gap, we develop a Multi-Agent-based Electrocatalyst Search Through Reasoning and Optimization (MAESTRO) framework in which multiple LLM agents collectively design active and stable single atom catalysts (SACs) for the oxygen reduction reaction (ORR) (**Figure 1**). Within this framework, agents iteratively propose structural modifications, reflect on outcomes and summarize design history, while rapid property evaluation is enabled by a machine learning force field (MLFF) serving as a surrogate for DFT[45].

Applying the MAESTRO framework, we demonstrate that the agents can autonomously formulate hypotheses regarding catalyst modification strategies and progressively improve both activity and stability through iterative reasoning. Notably, the framework identifies SACs that surpass the theoretical activity limit imposed by conventional scaling relations between reaction intermediates[46]. Detailed analysis reveals that this enhancement originates from the selective stabilization of a specific intermediate via hydrogen bonding, a mechanism previously reported in several studies. Importantly, this discovery does not emerge in the absence of in-context learning, providing strong evidence that the proposed framework enables agents to acquire new physics and design principles through accumulated history and iterative reasoning. These results highlight the MAESTRO framework as a promising optimization strategy for electrocatalyst discovery, capable of navigating complex design spaces and uncovering nontrivial relationships between structure and property beyond conventional discovery schemes.

## 2. Results

### 2.1 MAESTRO Framework

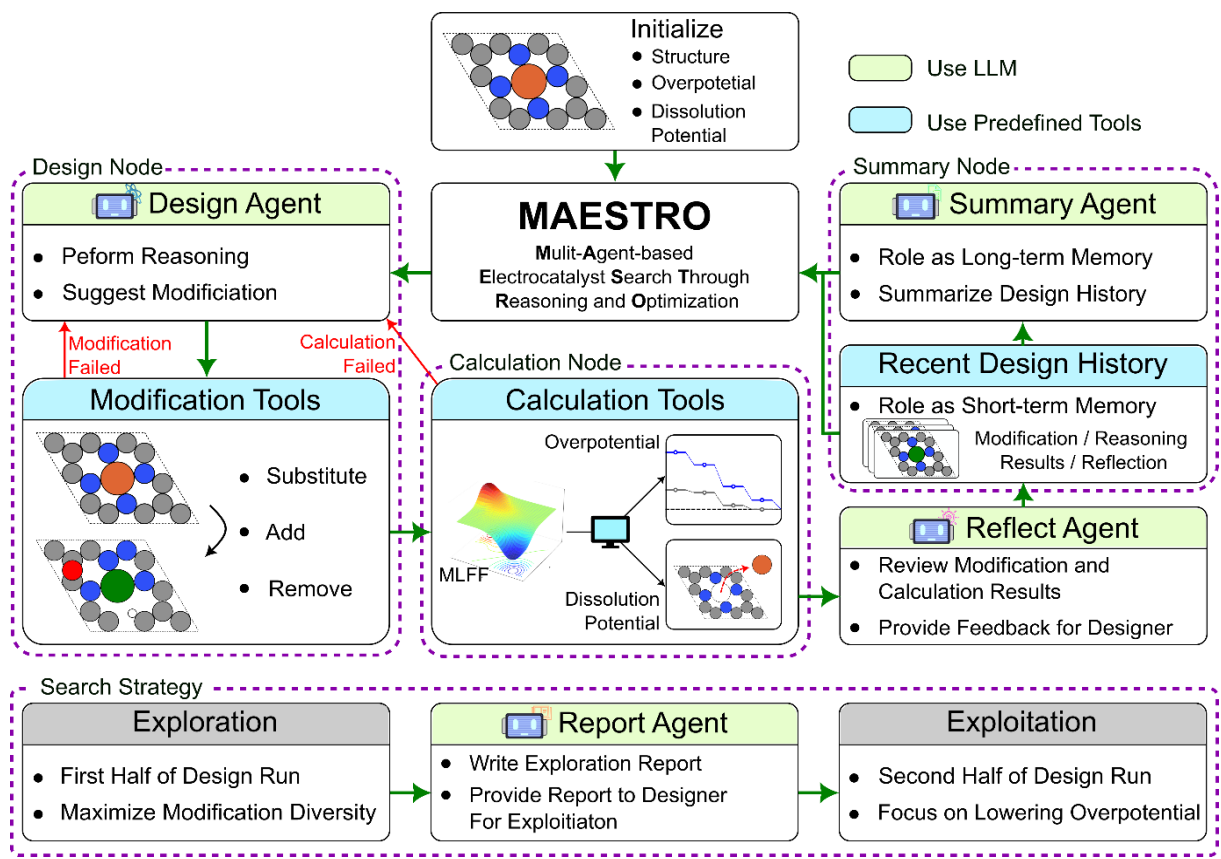

**Figure 1.** Overview of the MAESTRO framework and the exploration-exploitation search strategy. Starting from an initial catalyst structure and its associated properties, LLM-based agents and predefined tools iteratively perform reasoning, modification, evaluation, reflection and summarization according to their designated roles for a fixed number of cycles. The design run is divided into two phases, an exploration phase and an exploitation phase. Upon completion of the exploration phase, a summary report is generated by the agent.

The MAESTRO framework operates as an iterative design loop composed of four nodes, design, calculation, reflection and summary, managed by four agents, Design, Reflect, Summary and Exploration report, along with predefined tools. The primary objective of this loop is to minimize the ORR overpotential ($\eta$), which serves as a measure of the catalytic activity of the SAC, while simultaneously maximizing the dissolution potential ($U_{diss}$) to ensure electrochemical stability.

The design loop initiates by loading an initial SAC structure and evaluating its overpotential and dissolution potential using a MLFF surrogate model. The resulting geometric and catalytic data are then formatted and passed to the design agent within the design node. Leveraging structural images and formatted metadata, the design agent formulates a hypothesis to identify specific modifications, and the underlaying physical reasons, that could effectively tune binding energies to reduce the overpotential. The design agent is authorized to modify five distinct geometric components of the SAC: the center metal atom, first coordination shell, second shell, axial ligand and functional groups. The available modification types and options are summarized in **Table 1**.

**Table 1.** Geometric components of the SAC accessible to the design agent, including modification types and available chemical options. Specifically, axial ligands are introduced perpendicular to the surface, positioned beneath the binding site, while functional groups are added to the second coordination shell of the carbon support.

| Geometric Components | Modification Types | Atom / Molecule Options |
| --- | --- | --- |
| Center metal atom | Substitute | Pt, Pd, Ir, Ru, Fe, Co, Mn, Cu, Ni, Cr, V, Ti, Mo, Na, Ta, Ag, Au, Zn, Sn, Bi |
| First coordination shell | Substitute, Add, Remove | H, C, O, N, S |
| Second coordination shell | Substitute | C, P, B, S, N |
| Axial ligand | Substitute, Add, Remove | *O, *OH |
| Functional group | Substitute, Add, Remove | *COC, *COH, *CO |

The proposed modification is forwarded to the modification tool, which verifies whether the suggested change is applicable to the given SAC. If the modification is deemed unsuitable, the workflow returns to the design agent for a new proposal. This includes case in which the agent proposes invalid modifications, such as hallucinated modification type not included in the list or attempts to substitute non-existent elements. If accepted, the modification is applied, and the resulting structure is passed to the calculation node. Here, the overpotential and dissolution potential of the modified SAC are calculated using the MLFF. If the geometry optimization fails to converge within 100 optimization steps under a force threshold of 0.05 eV/Å, the failed structure and associated metadata are returned to the design agent. Upon successful convergence, the calculated properties are forwarded to the reflection node.

In the reflection node, the reflection agent evaluates the effectiveness of the proposed modification by comparing the catalytic activity and stability before and after the change. Based on this comparison, it provides feedback categorizing the modification as successful or unsuccessful. This feedback, along with the proposed modification, the underlying reasoning, and the calculated results, is passed to the summary node and formatted as design history. While the most recent history is stored in full, previous entries are condensed by the summary agent to maintain context efficiency (**Figure S2**). Finally, the accumulated design history, the modified SAC structure and its performance metrics are fed back to the design agent to initiate the next iteration.

To ensure sufficiently broad exploration, the agents are guided by an exploration-exploitation strategy. During the first half of the process, the exploration phase, the primary objective is to expand the design space rather than minimize the overpotential. Accordingly, the design agent is instructed to structurally diverse and distinct types of modifications without prioritizing improvements in overpotential or dissolution potential. The reflect agent provides feedback which prioritizes the discovery of unique structural configurations.

At the midpoint of the design agent, the workflow transitions to the exploitation phase, where optimization proceeds under the original objective of minimizing overpotential while maintaining stability. During this transition, an exploration report agent generates a one to two page report summarizing the modifications explored and their corresponding effects (**Figure S6**). This report distills actionable insights to guide the subsequent optimization. The exploration report serves as a persistent reference for the design agent throughout the remainder of the loop.

In this study, the primary results were obtained using GPT-4.1-mini as the LLM[29], Universal Models for Atoms (UMA) as the MLFF surrogate model[47], and $FeN_4$ as the initial catalyst structure. However, the framework is modular and additional experiments using alternative models and starting materials were also conducted. Detailed agent prompts, catalyst specifications and implementation details are provided in **Supplementary Note A**.

**2.2 Validation of MLFF and LLM components**

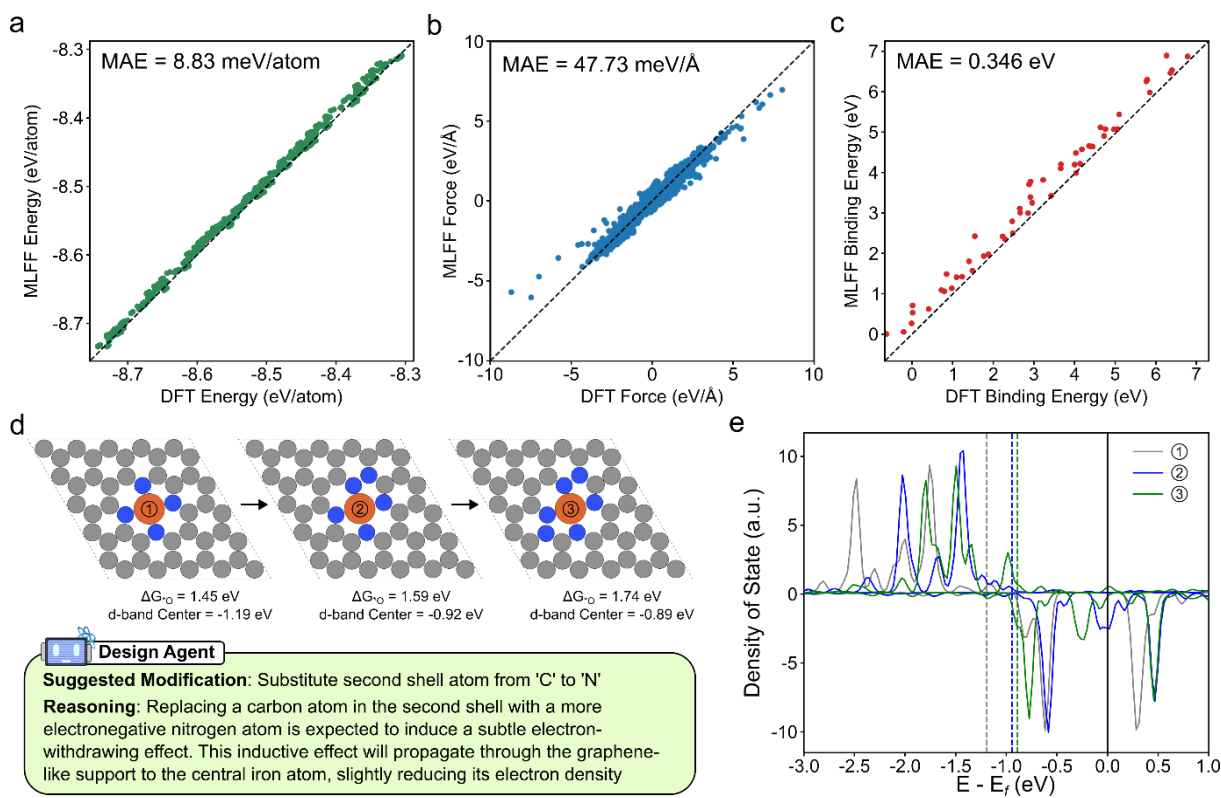

**Figure 2.** Pre-validation of MLFF and LLM components. (a-c) Parity plots comparing DFT-calculation and MLFF-prediction results for (a) energies per atom, (b) atomic forces and (c) binding energies. (d) An example of the reasoning and modification proposed by the design agent, together with the resulting structure and the changes in Gibbs free energy and d-band center induced by the suggested modification. (e) Corresponding changes in the electronic density of states (DOS) following the modification. Dashed lines indicate the d-band centers of the metal atoms, which shift toward positive values (closer to the Fermi level) upon the substitution of a second-shell carbon atom with nitrogen.

Before deploying the MAESTRO framework, we conducted a pre-validation study to ensure the reliability of both the tools and the LLM for the SAC system. Since the UMA employed as the MLFF in this study was not trained on SAC data, and no publicly available SAC dataset existed for meaningful fine-tuning, we constructed a custom SAC dataset to validate the extrapolation capabilities of UMA. The performance of the pre-trained UMA was evaluated using this out-of-distribution (OOD) dataset, which contains DFT calculated energies, forces and binding energies for various structures and their adsorbed intermediates. As shown in **Figures 2a** and **2b**, UMA with 'OC20' domain showed Mean Absolute Errors (MAEs) for DFT energies per atom and atomic forces of 8.83 meV/atom and 47.73 meV/Å, respectively. These values are comparable to the MAEs reported by Wood et al for OOD catalysts[47] (**Figure S3**). Furthermore, 0.346 eV MAE for binding energy was deemed acceptable, especially considering the complete absence of SACs in the pre-training set and the consistent directional bias of the prediction error. This systematic behavior indicates that UMA can reliably capture relative changes in binding strength, making it a suitable surrogate model for DFT within the design loop. Detailed information regarding the pre-validation datasets and MLFF performance is provided in **Supplementary Note B**.

Regarding the LLM, we validated whether the modification proposed by the model could effectively steer binding energies in the intended direction and whether the underlying reasoning was physically sound. The target was to achieve a binding energy of 2.46 eV for the *O intermediate. As shown in **Figure 2d**, the LLM successfully suggested a modification that shifted the *O binding energy from the initial structure toward the desired target. Specifically, the LLM hypothesized that N substitution would decrease the electron density of the center metal atom. This hypothesis was confirmed by density of states (DOS) calculations, which showed a clear shift of electronic density, quantified by the reduction of d-band center position, of the Fe center after substitution[48, 49]. While the correspondence between the intended and realized binding energy changes and their associated reasoning was not perfect in every case, it was correct in the majority of instances. Representative examples of unsuccessful suggestions are provided in **Figure S5**.

These pre-validation results demonstrate that UMA can function as a practical surrogate for DFT in evaluating the stability and catalytic activity of SACs for ORR without additional fine-tuning. Moreover, the LLM demonstrates a meaningful understanding of structure-property relationships of the SACs, enabling it to propose chemically interpretable modifications that systematically guide the energies toward target values.

**2.3 Performance Evaluation of Overall Framework**

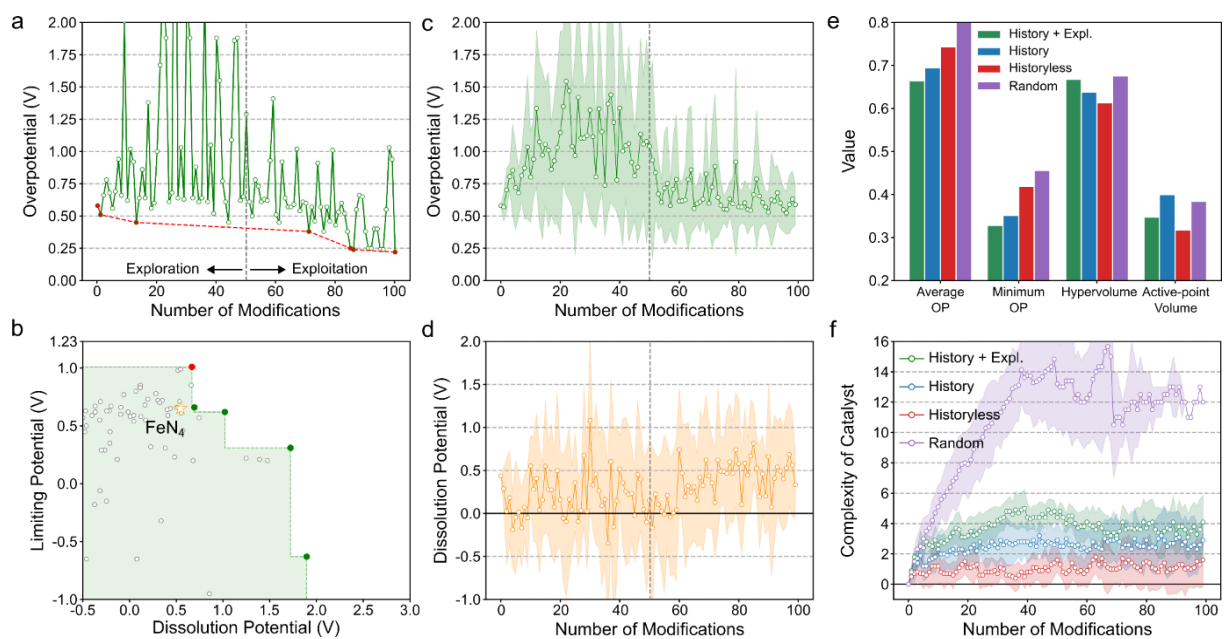

**Figure 3.** Performance and behavior of the MAESTRO framework across different strategies. (a) Representative example of the ORR overpotential change during a design run, illustrating the transition from the exploration phase to the exploitation after the 50[th] modification. The red dashed line denotes the running minimum overpotential. (b) Corresponding Pareto front obtained from the same run, where the red point highlighting the most active catalyst discovered. (c) Evolution of overpotential and (d) dissolution potential, averaged over 10 independent design runs using *"history + exploration"* strategy. (e) Comparison of overall design performance across different strategies. (f) Evolution of catalyst complexity during the design run for each strategy, where higher value indicates a larger number of modifications relative to the initial SAC structure.

The objective of the design run is to minimize the ORR overpotential while simultaneously preserving or increasing the dissolution potential. To systematically evaluate the performance of our framework, we employed four metrics.

1) Average overpotential: The mean ORR overpotential obtained across all design runs, excluding the exploration phase. For strategies without an explicit exploration phase, this was computed over the latter half of each design run.

2) Minimum overpotential: The average of the lowest overpotential achieved during each individual design run.

3) Hypervolume: The average volume of the Pareto front, representing the joint optimization space spanned by limiting potential and dissolution potential. A larger volume indicates superior discovery performance in terms of both activity and stability.

4) Active-point volume: The average Pareto volume of the catalyst with the lowest

overpotential, representing the combined activity and stability of the most promising candidate discovered.

To benchmark the performance of the proposed "*history + exploration*" strategy, which introduces an exploration phase to prioritize a global search before exploitation, we defined three baseline strategies: 1) The *history* strategy performs only exploitation without a preceding exploration phase, thereby isolating the effect of broadening the design space before focused optimization. 2) The *historyless* strategy relies solely on the LLM's background knowledge and does not incorporate information from previous modification steps, highlighting the impact of in-context learning from accumulated design history. 3) The *random* strategy applies purely random modifications without LLM guidance, serving as a lower-bound reference.

**Figure 3a** and **3b** present a representative design run, illustrating the gradual changes of overpotential and the corresponding Pareto front of the discovered catalysts. During the exploration phase, the overpotential exhibits large fluctuations as the framework broadly samples the design space. The exploratory behavior was further confirmed by the higher number of unique modifications performed during this phase, compared not only to the subsequent exploitation phase, but also to design runs of other strategies without exploration (**Figure S7**). In contrast, during the exploitation phase, the overpotential converges toward a narrow range, indicating focused optimization. The progression of the minimum overpotential, denoted by the red line and markers, reaches its optimum during this phase. The Pareto plot highlights the intrinsic trade-off between activity and stability. Although the most active catalyst identified in this run shows slightly lower stability, it still surpasses the reference $FeN_4$ catalyst in both metrics.

**Figure 3c** and **3d** summarize the average results over 10 independent design runs. Both graphs reveal a clear trend, where the overpotential gradually decreases while the dissolution potential increases following the transition to the exploitation phase (**Figure S8**). These observations demonstrate that the MAESTRO framework can effectively identify highly active SACs while maintaining or even improving their electrochemical stability. The origin of this enhanced discovery performance lies in the expansion of the accessible design space through the exploration phase and in the in-context learning enabled by the design history (**Figure 3e**). Our strategy exhibits the best performance in terms of both the average overpotential and the minimum overpotential. Notably, even the "*history*" strategy consistently outperforms the "*historyless*" approach in average overpotential, underscoring the importance of leveraging accumulated design history.

In contrast, the hypervolume and active-point volume metrics show comparable values across the different strategies. This behavior arises because these metrics can be disproportionately influenced by high stability values, which are generally easier to achieve than low overpotentials. Therefore, maintaining Pareto volume metrics comparable to other strategies while simultaneously achieving lower overpotentials provides strong evidence that the framework effectively mitigates the activity-stability trade-off. Finally, the moderate level of catalyst complexity, defined by the number of modifications applied to the initial structure, indicates that our strategy maintains a balanced search behavior (**Figure 3f**). The design agent neither diverges excessively into chemically unrealistic structures nor becomes trapped in a narrow, suboptimal region of the design space.

## 2.4 Impact of In-context Learning

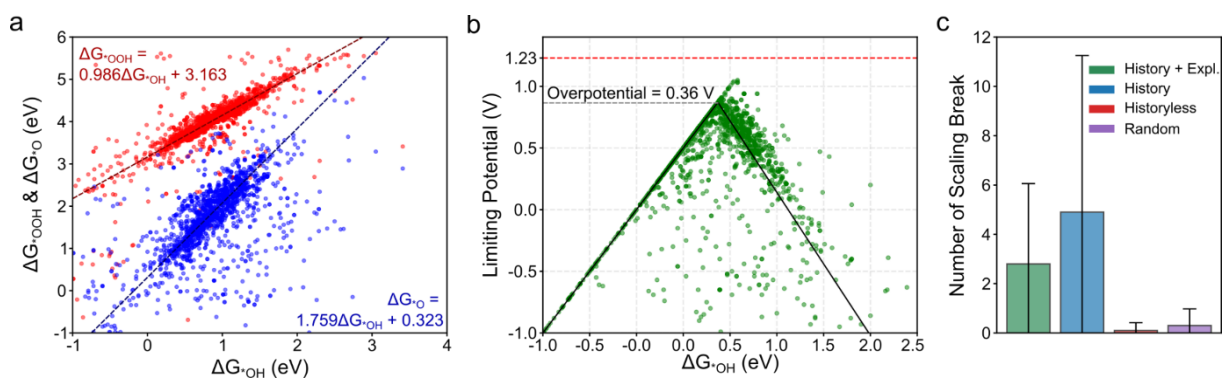

**Figure 4.** (a) Linear scaling relations among $\Delta G_{*OH}$, $\Delta G_{*O}$ and $\Delta G_{*OOH}$. (b) Corresponding volcano plot using $\Delta G_{*OH}$ as the activity descriptor. The red dashed line indicates the theoretical lower limit for overpotential 0.36 V imposed by the scaling relations. (c) Frequency of scaling relation violations observed during the design loop, averaged over 10 independent design runs for each strategy.

To elucidate the origin behind the good observed performance within the MAESTRO framework, we investigated how in-context learning from design history influences the design process. We collected the binding energies of 2,017 unique catalysts from a total 4,000 structures generated across four discovery strategies, where each strategy was evaluated over 10 independent design runs consisting of 100 modification steps each. We then analyzed the linear scaling relations among the binding energies of the three ORR intermediates, *OOH, *O and *OH. This analysis confirmed the presence of well-defined linear scaling relations between

ΔG*$_O$ and ΔG*$_{OH}$, as well as between ΔG*$_{OOH}$ and ΔG*$_{OH}$ in the SAC system (**Figure 4a**). Furthermore, by deriving a volcano plot using ΔG*$_{OH}$ as the descriptor, we identified a theoretical minimum overpotential of about 0.36 V (**Figure 4b**). This value represents the lowest overpotential achievable by uniformly strengthening or weakening the binding energies of all intermediates while remaining on the scaling relations, without selectively modulating individual intermediates (**Figure S9**).

As we confirmed in **Figure 2**, the LLM possesses background chemical knowledge that enables it to propose modifications to control binding strengths as intended. Consistently, in **Figure 3**, the *historyless* strategy, which relies only on this background knowledge, exhibited a minimum overpotential exceeding 0.36 V. This observation indicates that strategies achieving overpotentials below 0.36 V must have been acquired through leveraging in-context learning from prior design steps. Specifically, such strategies learn to selectively tune the binding energy of specific intermediates, thereby breaking the inherent scaling relations rather than merely shifting all binding energies in tandem[50]. This effect can be observed in observing the frequency of achieving overpotentials below lower limit (**Figure 4c**). Across 10 design runs, both the *random* and *historyless* strategies achieved this value fewer than once on average, indicating that breaking the scaling relations in these cases occurs only sporadically by chance. In contrast, strategies with in-context learning from history broke these relations more than three times on average out of 100 modification steps. These results demonstrate that in-context learning from modification history plays a decisive role in overcoming fundamental scaling constraints (**Figure S10**), despite the higher token consumption compared to strategies without in-context learning (**Figure S11**).

**2.5 Revealing Catalyst Design Principles**

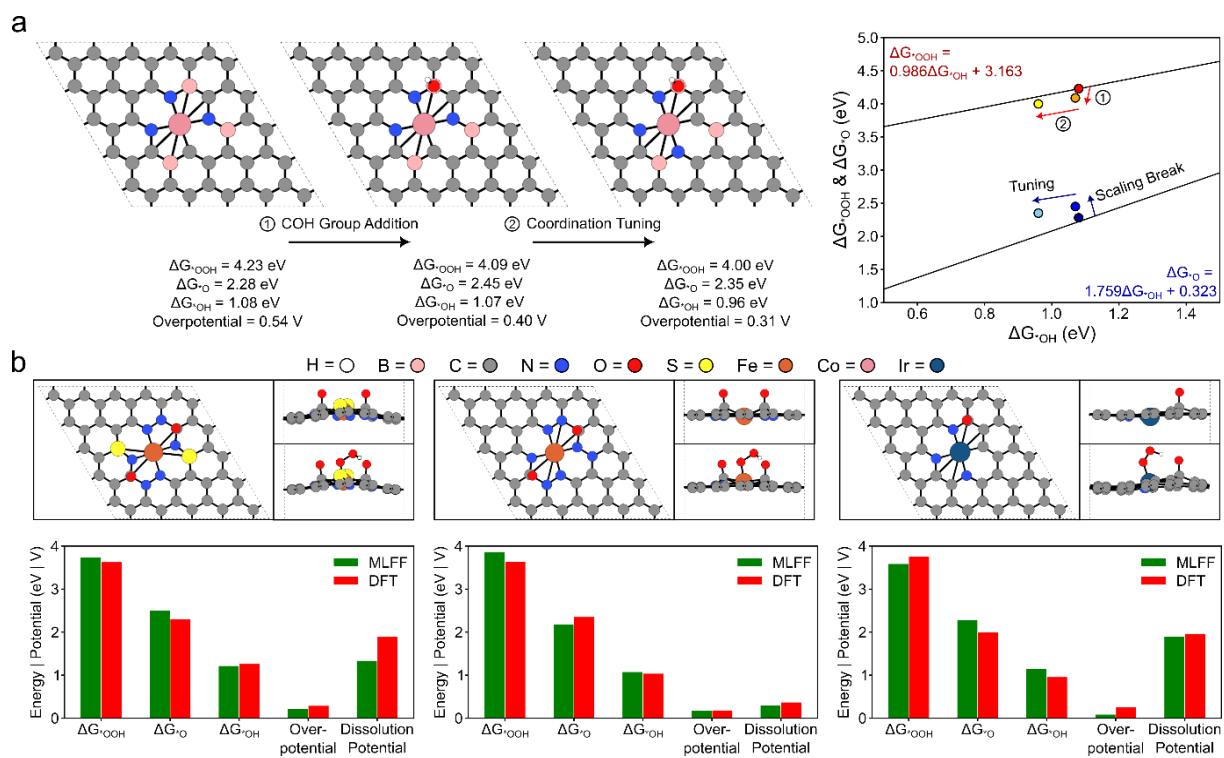

**Figure 5.** (a) Representative modification process to design promising catalysts by breaking scaling relations and lower limit of overpotential. Scaling break and overpotential optimization occur sequentially. (b) Representation of one of high-performance catalyst structures discovered during the design runs, together with a comparison of MLFF-predicted and DFT-calculated performance metrics. Data includes Gibbs free binding energies, overpotential ($\eta$) and dissolution potentials ($U_{diss}$) for each candidate. Other examples can be found in **Figure S12** and **S13**.

To elucidate the mechanism underlying the observed scaling relation breaking and the successful design of high-performance catalysts, we analyzed a representative designed catalyst in detail. **Figure 5a** shows an example of a catalyst achieving an overpotential of 0.31 V, surpassing the scaling-based lower limit of 0.36 V. During the design run, the design agent introduces surface oxygen functional groups (*COC or *COH). These surface oxygen species are observed to form H-bonds with the H atoms of *OH and *OOH intermediates, thereby stabilizing both species. As a result, only the $\Delta G_{*O}$ is selectively increased, leading to a break in the scaling relations between $\Delta G_{*OH}$ and $\Delta G_{*O}$. After inducing this scaling break, the agent retains the surface oxygen and subsequently explores first and second shell modifications, searching for configurations that further reduce the overpotential within the altered scaling landscape. Through this sequential process, the MAESTRO framework is able to design catalysts with overpotentials below 0.36 V. Importantly, this scaling break from selective H-bonds was confirmed not only at the MLFF but also through explicit DFT level. We performed DFT calculations on 11 catalysts with MLFF-predicted overpotentials below 0.36 V, selected from the minimum overpotential candidates obtained from each of the 10 design runs starting from FeN4 and CuN4. Among these, six unique catalysts have DFT-calculated overpotentials below 0.36 V, five of which were confirmed to feature surface oxygen forming H-bonds with ORR intermediates (**Figure 5b**).

The stabilization of intermediates via selective H-bonding, the resulting scaling break and the associated enhancement in catalytic performance have been reported in multiple prior studies[51-55]. Therefore, the framework does not introduce a fundamentally new design principle. However, as shown in **Figure 4c**, the agents without in-context learning lack the intrinsic ability to break scaling relations, indicating that these principles were not already present in their background knowledge, but were gained from experimentation. Indeed, through this exploration-exploitation strategy and in-context learning enabled by the design loop, the agents successfully rediscovered a valid design principle that existed outside their prior knowledge[56]. This result demonstrates that the MAESTRO framework has the capacity not only to optimize catalysts but also to potentially uncover novel catalyst design principles beyond those explicitly provided by human inputs.

## 3. Discussion

### 3.1 LLM, Starting State and Hyperparameter Tuning

We have demonstrated that the MAESTRO framework operates successfully using GPT-4.1-mini as the LLM and FeN$_4$ as the initial catalyst. We further examined how the choice of LLM, the starting SAC structure and key hyperparameters, including the LLM temperature and the number of recent design histories retained in short-term memory, affect the overall performance.

First, when the LLM was replaced with GPT-5-mini while all other conditions were kept identical, the agents exhibited slightly more detailed reasoning and feedback. However, the overall performance remained comparable to that obtained with GPT-4.1-mini. This observation indicates that GPT-5-mini possesses background knowledge similar to that of GPT-4.1-mini and that, in both cases, the ability to overcome scaling relations must be acquired through in-context learning during the design loop rather than being directly encoded in the model (**Figure S14a**).

We also investigated the effect of the starting material by initializing design runs with SACs containing other central metal atoms. In all cases, the average minimum overpotential achieved fell below 0.36 V, the limit dictated by the volcano plot. Moreover, each design run broke the scaling relations more than twice on average, demonstrating that our strategy is robust with respect to the choice of the initial metal center (**Figure S14b**). Among the starting structures, the design run initiated form $PtN_4$ exhibited the best overall performance, except for the average overpotential. This behavior can be attributed to the intrinsically high initial overpotential of $PtN_4$, which allows for broader exploration to the energy landscape during the exploration phase.

Finally, we evaluated the sensitivity of the framework to the temperature settings by performing 10 design runs across a range from 0.2 to 1.8. The results indicate that, provided the temperature remains within a moderate range, the overall discovery performance is largely comparable to the default setting of 1.0. In contrast, extreme values, such as 0.2 or 1.8, lead to a noticeable performance degradation (**Figure S15**).

### 3.2 Limitations and Future Directions

The types of SAC modifications employed in this study focus on fine-tuning the local environment of the active site by element substitution, atom addition/removal, and the introduction of functional groups or ligands. Although this approach expands the accessible design space, catalysts identified through such localized modifications may pose challenges for experimental synthesis. In particular, experimentally reproducing an identical local atomic environment remains difficult in practice, even when the designed structures satisfy electrochemical stability criteria. As a result, direct experimental validation of some discovered catalysts may be non-trivial. In subsequent studies, the inclusion of better metrics for synthesizability or even synthesizability predictions will facilitate bridging the gap to experimental validation.

Furthermore, this study is limited to the discovery of ORR catalyst within the SAC system. This choice was made to establish a proof-of-concept for agent-based catalyst discovery, leveraging SACs as a platform in which the active site is spatially confined and structural complexity is relatively low. Despite these simplifications, the results demonstrate that the framework is capable of selectivity modulating the binding energies of multiple intermediates while simultaneously managing activity, stability and structural complexity. These capabilities suggest that the framework can be extended to more complex systems, such as dual atom catalysts (DACs) or to reactions like $CO_2$ reduction, where the independent control of intermediate energies is even more critical.[57]

## 4. Conclusions

In this work, we proposed the MAESTRO framework and demonstrated that iterative interactions among specialized agents within an optimization loop can progressively enhance both catalytic activity and stability. Notably, the catalysts identified by the framework break the theoretical lower limit of ORR overpotential imposed by the conventional scaling relations. These findings indicate that the accumulation of design history, combined with agent-driven reasoning and self-reflection, can reveal new physical principles not explicitly encoded in the LLM's background knowledge. Overall, our results suggest that the MAESTRO framework serves not only as an effective catalyst optimization tool but also as a means of autonomously generating novel chemical insights, significantly reducing the need for human intervention in the discovery of next-generation catalysts.

## 5. Methods

### 5.1 Large Language Models

A large language model (LLM) is a transformer-based autoregressive generative model trained on large-scale text corpora[58]. By learning statistical and semantic patterns in language, an LLM is capable of performing complex reasoning, planning and decision-making tasks that resemble human cognitive processes. In this study, unless otherwise specified, all LLM-based agents were implemented using OpenAI's GPT-4.1-mini model with default temperature and top-p settings. The detailed personas and system prompt templates assigned to each agent are provided in **Supplementary Note 1**.

### 5.2 Machine Learning Force Field

Machine learning force field (MLFF) models serve as surrogate models for density functional theory (DFT) calculations. They are trained on large DFT dataset to predict energy and atomic forces based on the geometry and stoichiometry of materials[45]. These models typically interpret materials as graphs, where atoms are treated as nodes characterized by elemental features and interatomic bonds are encoded as edges containing geometric information. In this study, we employed the Universal Models for Atoms (UMA) as the MLFF[47]. UMA is based on eSEN[59], an equivariant graph neural network that represents the atomic environment using spherical harmonic embedding and propagates information through multiple message passing layers. UMA is trained as a general-purpose model across diverse material domains and incorporates a Mixture of Linear Expert (MoLE) architecture to enhance flexibility and inference efficiency[60]. For geometry optimization and energy prediction of the SACs, we utilized the pretrained UMA model corresponding to the 'OC20' domain[61], which is specifically weighted toward heterogenous catalyst systems.

### 5.3 DFT Calculations

We performed DFT calculations using the Vienna Ab initio Simulation Package (VASP, version 5.4.4)[62, 63] to pre-validate MLFF performance and to evaluate SAC structures discovered by the catalyst design framework. To ensure consistency, the same DFT setting used to generate the UMA training data were adopted. The projected-augmented wave (PAW) pseudopotential method[64] was employed in conjunction with the generalized gradient approximation revised Perdew-Burke-Ernzerhof (GGA-RPBE) exchange-correlation functional[65]. All structures were fully relaxed until the total energy and atomic forces converged to within $10^{-4}$ eV and 0.05 eV/Å, for forces, respectively. A plane wave kinetic energy cutoff of 350 eV was applied. Monkhorst-Pack k-point mesh[66] was configured as (3 × 3 × 1).

### 5.4 Gibbs free energy, Overpotential and Dissolution Potential

Binding Gibbs free energies were evaluated using the computational hydrogen electrode (CHE) approach[67], in which the chemical potential of a proton-electron pair ($H^+ + e^-$) is referenced to ½$H_2$ (g) at 0 $V_{RHE}$ under standard conditions. Baded on this framework, the Gibbs free energies for the adsorption of *O, *OH and *OOH intermediates were computed as follows:

$$\Delta G_{O*} = E_{O*} - E_{slab} - E_{H_2O} - E_{H_2} + \Delta(ZPE + \int C_p dT - TS)$$

$$\Delta G_{OH*} = E_{OH*} - E_{slab} - E_{H_2O} - \frac{1}{2}E_{H_2} + \Delta(ZPE + \int C_p dT - TS)$$

$$\Delta G_{OOH*} = E_{OOH*} - E_{slab} + 2E_{H_2O} - \frac{3}{2}E_{H_2} + \Delta(ZPE + \int C_p dT - TS)$$

Here, $E_{*O}$, $E_{*OH}$ and $E_{*OOH}$ denoted the DFT total energies of the surface with the corresponding adsorbed intermediates, while $E_{H_2}$ and $E_{H_2O}$ are DFT energies of gas phase $H_2$ and $H_2O$ molecules, respectively. Zero-point energies (ZPE) corrections, enthalpic contribution ($\int C_p dT$), and entropic contribution terms (TS) were calculated using the Harmonic oscillator approximation for adsorbed species (*O, *OH, *OOH) and the Ideal gas approximations for gas molecules ($H_2$, $H_2O$), as implemented in Atomic Simulation Environment (ASE)[68]. The correction values are given in **Table S8**.

The theoretical overpotentials of ORR ($\eta^{ORR}$) was determined from Gibbs free energy changes of the four elementary proton-electron transfer steps in the associative ORR pathway:

$$\Delta G_1 = \Delta G_{*OOH} - \Delta G_{O_2} + eU$$

$$\Delta G_2 = \Delta G_{O*} - \Delta G_{*OOH} + eU$$

$$\Delta G_3 = \Delta G_{*OH} - \Delta G_{*O} + eU$$

$$\Delta G_4 = \Delta G_{H_2O} - \Delta G_{*OH} + eU$$

The limiting potential ($U_L$) is defined as the maximum potential at which all reaction steps become thermodynamically favorable ($\Delta G_i(U_L) \leq 0$ eV). The overpotential is the difference between the standard equilibrium potential for ORR (1.23 V) and the limiting potential, calculated as follows:

$$U_L = -\max[\Delta G_1, \Delta G_2, \Delta G_3, \Delta G_4]/e$$

$$\eta^{ORR} = 1.23 - U_L$$

The dissolution potential of the SAC ($U_d$) is an electrochemical stability, describing the tendency of the metal center to dissolve under ORR conditions. It is calculated as the difference between the binding energy of the metal atom to the carbon support ($E_b$) and the standard dissolution potential of the corresponding bulk metal ($U_0$) in aqueous solution (pH = 0), as follows:

$$E_b = E_{SAC-M} - E_{SAC} - \mu_M$$

$$U_d = U_0 - E_b / (e \times N_e)$$

where $E_{SAC-M}$, $E_{SAC}$ are the DFT total energies of the carbon support with and without the metal atom, respectively. $\mu_M$ is the chemical potential of the metal atom, $e$ is elementary charge and $N_e$ is the number of electrons involved in the metal dissolution process.

**Code Availability**

The code developed in this work and relevant information can be found in GitHub (https://github.com/ahrehd0506/Catalyst-Design-Agent)

**Supporting Information**

Details of prompt template, Details of pre-validation for MLFF and LLM, Example of exploration summary report and reasoning of the agents, Additional metrics for design framework, Metrics for strategies with various starting materials, parameter and LLM model.

**Acknowledgements**

D.H.M. acknowledges the support from Korea Institute for Advancement of Technology (KIAT) grant funded by the Ministry of Trade, Industry & Energy (MOTIE), Korea Government (RS-2024-00436106, Human Resource Development Program for Industrial Innovation). S.B.

acknowledges the support from the National Research Foundation of Korea (NRF) grants funded by the Korea government (MSIT and MOE) (RS-2024-00448287, RS-2025-16063688, RS-2025-00513832, and RS-2025-02214715), and the generous supercomputing time provided by the Korea Institute of Science and Technology Information (KISTI). G.H. acknowledges support from the U.S. National Science Foundation under Grant # CBET-2442223. This work used NCSA Delta CPU at University of Illinois Urbana-Champaign through allocation MAT250081 from the Advanced Cyberinfrastructure Coordination Ecosystem: Services & Support (ACCESS) program, which is supported by U.S. National Science Foundation grants #2138259, #2138286, #2138307, #2137603, and #2138296.

# Supplementary Information

**Supplementary Note A. Details of Prompt Template**

**A.1 System Prompt**

Four discovery strategies were implemented within Multi-Agent-based Electrocatalyst Search Through Reasoning and Optimization (MAESTRO) framework: *history + exploration*, *history*, *historyless* and *random*. Except *random* strategy, which relies on stochastic modification selection, other strategies utilized LLM-based agents. For the LLM-based strategies, both the design and reflect agents operate under strategy specific configurations. In particular, their system prompts, which govern overall behavior throughout the entire design loop, differ depending on the selected strategy (**Table S1** and **S2**). The summary and exploration report agent do not employ strategy specific prompt variations at the system or input prompt level. Instead, their participation in the framework is determined solely by the chose strategy. Specifically, the summary agent is deactivated for strategies that do not utilize design history, while the exploration report agent is deactivated for strategies that do not include exploration phase (**Table S3**).

**A.2 Output Format**

The design and reflect agents employ fixed and structured output formats to ensure robust interaction throughout the design loop. These output formats remain identical across all discovery strategies. If an agent generates a response that deviates from the prescribed format, for example due to hallucinated or malformed outputs, the response is returned to the agent together with explicit feedback describing the formatting error. When an agent repeatedly produces invalid outputs beyond a predefined threshold, the corresponding design run is automatically terminated.

The output of the design agent consists of the selected modification type, the associated parameters and the reasoning underlying the proposed change. The reflect agent produces structured feedback that evaluates both the proposed modification and the resulting performance. Based on this reflection, the agent recommends the next catalyst to be modified by selecting from a set of recently modified catalysts stored in the design history. This recommendation mechanism functions as an "undo" operation, allowing the framework to revert to a previous design state when the modified catalyst diverges excessively from the target region or becomes trapped in repetitive modification cycles.

**A.3 Scientific Rule**

To ensure consistent physical interpretation across agents, all agents were instructed in the electrochemical convection for binding Gibbs free energy. Unlike general chemical context, where high energy typically indicates strong interactions, binding in electrochemistry is characterized by stronger binding at lower (more negative) energy value. Because LLM background knowledge is largely dominated by general chemistry conventions, omission of this distinction can lead to systematic reasoning errors. To mitigate this issue, we incorporated few-shot examples and terminology to enforce the correct interpretation of binding energy (**Table S4**)

**Table S1.** System prompt templets of design agent

| Design Agent System Prompt Base |
|---|
| You are an expert Computational Chemist Designer specialized in Single Atom Catalysts (SACs).<br>{strategy}<br><br>{scientific_rules}<br><br>{output_format} |
| **{strategy}** |
| *history + hxploration* |
| Your task:<br>- To propose one well-reasoned modification for the current catalyst to reach the target property.<br>- To keep the catalyst stable. The higher dissolution potential indicates higher stability.<br>- To keep the catalyst from being too complex.<br><br>You will be given:<br>- The current catalyst description<br>- Target type/value or target ranges<br>- Recent modification history<br>- Self-reflection of previous modification<br>- A discovery strategy (exploration or exploitation)<br><br>You MUST:<br>- Propose exactly ONE modification in the 'modifications' list.<br>- Write a hypothesis (8 to 11 sentences) explaining WHY this modification should move the system toward the target.<br>- If strategy is 'exploration', DO NOT use same modifications in history.<br>- If strategy is 'exploitation', MUST choose modification based on history. |
| *historyless* |
| Your task:<br>- To propose one well-reasoned modification for the current catalyst to reach the target property.<br>- To keep the catalyst stable. The higher dissolution potential indicates higher stability.<br>- To keep the catalyst from being too complex.<br><br>You will be given:<br>- The current catalyst description<br>- Target type/value or target ranges<br><br>You MUST:<br>- Propose exactly ONE modification in the 'modifications' list.<br>- Write a hypothesis (8 to 11 sentences) explaining WHY this modification should move the system toward the target. |
| Others |
| Your task:<br>- To propose one well-reasoned modification for the current catalyst to reach the target property.<br>- To keep the catalyst stable. The higher dissolution potential indicates higher stability.<br>- To keep the catalyst from being too complex.<br><br>You will be given:<br>- The current catalyst description<br>- Target type/value or target ranges<br>- Recent modification history<br>- Self-reflection of previous modification<br><br>You MUST:<br>- Propose exactly ONE modification in the 'modifications' list. |
| - Write a hypothesis (8 to 11 sentences) explaining WHY this modification should move the system toward the target. |
| **{output format}** |
| {"modifications": [{<br>    "modification_type": "$TYPE",<br>    "parameters": ["$PROPERTY_1", "$PROPERTY_2"],<br>    "reasoning": "$HYPOTHESIS",<br>}]}<br><br>$HYPOTHESIS: Your scientific reasoning. Explain why chosen modification will move the SAC toward the target value.<br>$TYPE: One of the allowed modification types.<br>$PROPERTY_1: The first parameter (e.g., element to remove). If the type does not require a parameter, use "None".<br>$PROPERTY_2: The second parameter (e.g., new element). If the type only requires one parameter, use "None". |

**Table S2.** System prompt templets of reflect agent

| |
|---|
| **Reflect Agent System Prompt Base** |
| You are an expert Computational Chemist Criticism specialized in Single Atom Catalysts (SACs).<br>{strategy}<br><br>Your task<br>1) Reflection (<= 7 sentences):<br>- State whether the modification moved the catalyst toward the target or away from it,<br>- If the result looks unphysical or failed, provide reflection on the modification why it was failed.<br><br>2) Choose the NEXT starting catalyst (undo mechanism):<br>You MUST choose exactly one source:<br>- next_catalyst_type = 'recent' > continue from one of RECENT candidates<br>- next_catalyst_type = 'best' > revert to BEST (global best so far)<br><br>3) If you choose NEXT starting catalyst as 'recent', choose index[0..N-1] (0=oldest, N-1=most recent)<br><br>Decision guidelines:<br>- Choose the MOST RECENT catalyst (RECENT[N-1]) if it is physically reasonable and not worse than other options.<br>- Choose an earlier RECENT index if the most recent result is clearly unphysical, overly complex, or significantly further from the target than a previous recent catalyst.<br>- Choose 'best' if:<br>  (a) RECENT candidates show repeated unphysical/divergent behavior (ex, all target value of recent candidates is far from the best), OR<br>  (b) BEST is clearly the most reliable and closest-to-target state among BEST vs RECENT, especially when the most recent step degraded the result.<br><br>{output_format}<br><br>{scientific_rules} |
| **{strategy}** |
| *history + exploration* |
| You will be given:<br>- The target type/value<br>- The modification + hypothesis proposed by the DesignAgent<br>- Catalyst description and evaluated values before/after the modification<br>- Optional images of optimized structures,<br>- Current catalyst complexity (higher = more complex),<br>- A list of the N most recent catalysts (RECENT, chronological: 0=oldest, N-1=most recent),<br>- The global best catalyst so far (BEST) with its property and complexity.<br>- A discovery strategy (exploration or exploitation).<br><br>Strategy-dependent decision guidelines<br>1) Exploration: |
| - Prefer continuing from RECENT catalyst that has potential to expand catalyst space and avoid choosing catalysts already used.<br>- Never use BEST.<br><br>2) Exploitation<br>- Prefer reverting to BEST or RECENT catalyst with good performance when the most recent step degraded performance or increased complexity without benefit.<br>- Continue from RECENT only if the new state is reliable and comparable to BEST (or clearly improving toward target). |
| Others |
| You will be given:<br>- The target type/value<br>- The modification + hypothesis proposed by the DesignAgent<br>- Catalyst description and evaluated values before/after the modification<br>- Optional images of optimized structures,<br>- Current catalyst complexity (higher = more complex),<br>- A list of the N most recent catalysts (RECENT, chronological: 0=oldest, N-1=most recent),<br>- The global best catalyst so far (BEST) with its property and complexity. |
| {output format} |
| {<br>  "reflection": "<your assessment of the current modification (<=6 sentences)>",<br>  "next_catalyst_type": <'recent' or 'best'><br>  "next_catalyst_index": <integer index in [0, N-1]>,<br>  "next_catalyst_reason": "<4~5 sentences explaining why this catalyst is the best starting point for the next iteration and why didn't choose another option>"<br>} |

**Table S3.** System prompt templets of summary and exploration report agent

| |
|---|
| **Summary Agent System Prompt Base** |
| You are a Historian of a Computational Chemist experiment specialized in Single Atom Catalysts (SACs)<br>You will be given a list of modification steps (modification, results) performed by catalyst design agent.<br><br>Your task is to summarize the progress into a concise paragraph.<br>Present the summary while keeping the following points in mind.<br>- Which modification induced an increase or decrease in Gibbs free energy (Delta G).<br>- Which modification was successful or unsuccessful in achieving the target.<br>- Which modification was the most critical.<br><br>Keep it under 200 words. This summary will be read by the Designer to decide the next step.<br><br>{scientific_rules} |
| **Exploration Report Agent System Prompt Base** |
| You are ReportAgent, an expert Computational Chemist analyst specialized in Single Atom Catalysts (SACs).<br>You are invoked once after iterations of EXPLORATION and immediately before switching to EXPLOITATION.<br><br>You will be given:<br>- The full exploration history: each iteration's starting catalyst, applied modification (type + parameters), DesignAgent hypothesis, evaluator results (before/after values), ReflectionAgent assessment, and complexity.<br><br>Your task:<br>Write a concise, 1-page report that summarizes exploration outcomes in a way that directly improves exploitation decisions. This is not a narrative; it is a decision-ready technical brief.<br><br>{scientific_rules}<br><br>Report Requirements<br>- Length: ~1 page (roughly 350~600 words). Be compact, information-dense, and structured.<br>- Use clear section headers and bullet points. |

> - Only include information supported by the provided history/results; do not invent new experiments.
> - When you mention a claim (e.g., "ligand OH lowers *OOH"), back it with at least one concrete example from history.
> - If some trend is weak or inconsistent, state that explicitly.
>
> Must-Focus Topics
> 1) Modification -> Outcome Mapping
> Identify which modification types/parameters, when applied to which catalyst contexts, produced:
> - Improvement toward target vs degradation
> - Physically reasonable vs unphysical/distorted outcomes
> Summarize as "pattern statements" + 2~5 concrete example bullets each.
>
> 2) Selective Adsorbate Tuning Patterns
> Extract modification patterns that selectively tune one adsorbate's Delta G more than others.
> Examples of the style you must produce:
> - Keeps Delta G(*O) and Delta G(*OH) roughly stable while shifting Delta G(*OOH) slightly downward/upward
> - Primarily weakens/strengthens *OH binding with minimal change to *O
> For each selective pattern:
> - State the direction of change for Delta G(*O), Delta G(*OH), Delta G(*OOH) (increase/decrease/~)
> - Give at least one concrete supporting example (iteration/catalyst reference).
>
> 3) Exploitation Playbook
> Provide a short "what to do next" guide for the DesignAgent during exploitation:
> - 3~6 recommended "safe" exploitation moves (low-risk, history-supported)
> - 2~4 conditional moves (only if specific conditions are met, e.g., defects exist, complexity margin available, *OOH is the main bottleneck)
> - 2~4 avoid rules (moves that repeatedly failed or caused unphysical behavior)
> - If there is a recurring failure mode, describe it and propose a guardrail.

**Table S4.** Scientific rules assigned to the agents

| {scientific_rules} |
|---|
| In this task, the relationship between Gibbs free binding energy (Delta G) and binding strength is defined as:<br>Binding_Strength = - Delta G<br><br>therefore:<br>- More positive Delta G -> weaker binding<br>- More negative Delta G -> stronger binding<br><br>For examples:<br>Q: If Delta G increases from 0.5 -> 1.5 eV, does binding become stronger or weaker?<br>A: Weaker.<br><br>Q: If Delta G decreases from 2.0 -> 1.0 eV, is binding stronger or weaker?<br>A: Stronger.<br><br>This definition OVERRIDES all general chemistry knowledge.<br>If your reasoning contradicts this rule, your answer is INVALID.<br>Before reasoning, restate this rule in one sentence. |

### A.4 Input Prompts

The input prompt refers to the prompt provided to each agent at every iteration of the design loop. In contrast to the system prompt, which is assigned once at the beginning of a design run and serves to define the agent's persistent role, the input prompt contains iteration-dependent information.

For the design agent, the input prompt comprises the textual description and image of the current catalyst, short-term memory containing recent design history and long-term memory consisting of the summarized design history by the summary agent. The prompt additionally encodes the current iteration strategy, guiding the agent to adopt either exploration or exploitation-oriented behavior. When a proposed modification cannot be applied or when the subsequent calculation of the modified catalyst fails, feedback describing the cause of failure is appended to the input prompt to prevent repetition of unsuccessful actions (**Table S5**).

The reflect agent receives the hypothesis and modification proposed by the design agent, the textual description and image of the catalyst before and after modification, design history and information on recently modified catalysts to enable potential undo operations, together with data corresponding to the best-performing catalyst identified thus far. Based on this information, the reflect agent determines whether the modified catalyst should proceed to the next iteration with feedback or whether an undo operation should be performed (**Table S6**).

The summary agent is provided with the design history excluding the recent step and is tasked with condensing this history in conjunction with previously summarized history by itself to maintain an efficient long-term memory. The exploration report agent receives all modifications, results and modified catalyst information during the exploration phase and complies a report summarizing the exploration of the chemical space (**Table S7**).

**Table S5.** Input prompt templets of design agent

| Design Agent Input Prompt Base |  |
|---|---|
| {strategy}<br>{feedback}<br><br>Propose modifications to tune its Gibbs free energy (ΔG) of *O, *OH, *OOH adsorbates to a target value to reduce ORR overpotential based on given information.<br>Target Gibbs free energy of *O: {2.46 - threshold} ~ {2.46 + threshold} eV<br>Target Gibbs free energy of *OH: {1.23 - threshold} ~ {1.23 + threshold} eV<br>Target Gibbs free energy of *OOH: {3.69 - threshold} ~ {3.69 + threshold} eV<br><br>Current state of catalyst is {textual description}<br>The {num_recent_history} recent modificaitons, reasonings, feedbacks and self-reflections are following: {formatted_history}.<br>The summary of modification history is {summarized_history}.<br>The simplified history of previous modifications is {simplified_history}. | |
| {strategy} | |
| Exploration Phase | Exploitation Phase |
| STRATEGY: EXPLORATION<br>Your primary objective is to explore uncertain but plausible regions of the catalyst space<br>Rule:<br>- You MUST not suggest the modifications alreday used in history<br>- Just make sure the complexity doesn't exceed the maximum, do not care about it below that.<br><br>Below is the list of modifications that alreday used: {modification_list} | STRATEGY: EXPLOITATION<br>Your primary objective is to explore uncertain but plausible regions of the catalyst space<br>Rule:<br>- Choose proper modifications to reach the target by referring to the Gibbs free energy change in the previous history and report.<br>- Try to maintain complexity of catalyst moderate<br>- Never suggest the failed combination of the modification + current catalysts already used in history.<br><br>Below is the report that summarized the exploration phase: {exploration_report} |

| You must still aim toward the target, but information gain has priority over immediate best performance. | Performance toward the target has priority over novelty. |
|---|---|
| {feedback} | |
| If recent modification succeeds | If recent modification fails |
| - | Your recent modifications {previous_modifications} are failed.<br>The reason of the most recent failure is {failed_reason}.<br>Please re-propose the modification based on the given format and information.<br>NEVER suggest same modification with previous failed modifications |

**Table S6.** Input prompt templets of reflect agent

| Reflect Agent Input Prompt Base |
|---|
| {strategy}<br><br>Design agent suggested following hypothesis and modification.<br>Hypothesis: {reasoning}<br>Modification: {modification}<br><br>After completing the modification, we obtained the following catalyst<br>Before modification: {previous_catalyst_textual_description}<br>After modification: {current_catalyst_textual_description}<br><br>Please write a brief post-action reflection on the modification in less than five sentences, \\<br>explaining how successful it was in achieving<br>{2.46 - threshold} ~ {2.46 + threshold} eV for Gibbs free energy of *O<br>{1.23 - threshold} ~ {1.23 + threshold} for Gibbs free energy of *OH and<br>{3.69 - threshold} ~ {3.69 + threshold} for Gibbs free energy of *OOH,<br>and the reasons for its success or failure.\\<br><br>The {num_recent_history} recent modificaitons, reasonings and feedbacks is following:<br>{formatted_history}.<br>The summary of history is {summarized_history}<br>The simplified history of previous modifications is {simplified_history}.<br>Recommended maximum value of complexity is {max_complexity}.<br><br>Global best catalyst found so far is {best_catalyst_textual_description}.<br>Recent catalyst list is following (oldest to newest):<br>{recent_catalysts_list}<br>If all target value of recent catalysts is fall apart from the best, choose 'best' |
| {strategy} |

| Exploration Phase | Exploitation Phase |
|---|---|
| STRATEGY: EXPLORATION<br>Your priority is information gain and coverage of the catalyst space.<br>- Prefer continuing from RECENT catalyst that has potential to expand catalyst space and avoid choose catalysts already used.<br>- MUST explore as many catalyst + modification possibilities as possible. | STRATEGY: EXPLOITATION<br>Your priority is fast convergence toward the target with reliable, low-risk steps.<br>- Recommends you judge to be the most suitable catalyst for the designer to achieve the target..<br>- Balance the use of RECENT and BEST. DO NOT stick to the BEST and select catalysts with the potential to achieve new BEST. |

**Table S7.** Input prompt templets of summary and exploration report agent.

| Summary Agent Input Prompt Base |
|---|
| Following is recent summarized history by yourself: |
| {summarized_history}<br>And following is recent detailed history:<br>{formatted_history}<br><br>Please summarize history based on given summarized and detailed history\ |
| **Exploration Report Agent Input Prompt Base** |
| Following is detailed history:<br>{formatted_history}<br><br>Please write the report based on history |

## A.5 Formatting

To communicate catalyst information to LLM-based agents, the catalyst structure must be expressed in a form compatible with natural language. However, accurately describing geometric details such as interatomic bonding and atomic coordinate using text alone remains challenging. Although several approaches for representing materials in natural language have been proposed[1], this study focuses on the relatively simple single atom catalyst (SAC) system. Accordingly, we designed a textual description that encodes only the key structural features characteristic of SACs and energy-related properties as shown in **Figure S1a**. In addition, all LLMs employed in this work are multimodal and capable of processing image inputs. Since two orthogonal perspectives are sufficient to capture the full geometry of SACs, both top view and side view images of each catalyst and its binding configurations were provided to the agents alongside the textual description (**Figure S1b**).

The design history was also formed to facilitate efficient interpretation by the LLMs. Two distinct history formatting methods were employed. The first format retains comprehensive information, including the modification type, reasoning from the design agent, and the feedback from the reflect agent. This formatted history serves as short-term memory. The second format is a condensed representation that records only the applied modification and their corresponding energy changes, excluding agent intervention. This simplified history functions as long-term memory preserving only objective state transitions.

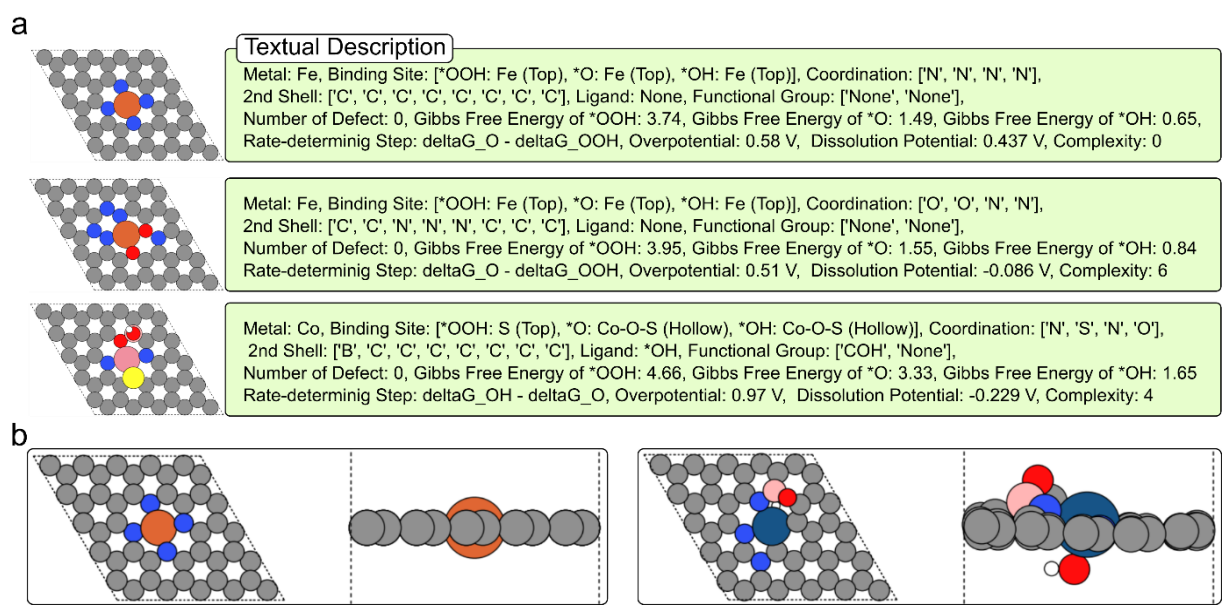

**Figure S1.** (a) Examples of textual description of SAC. (b) Examples of SAC image provided to the agents.

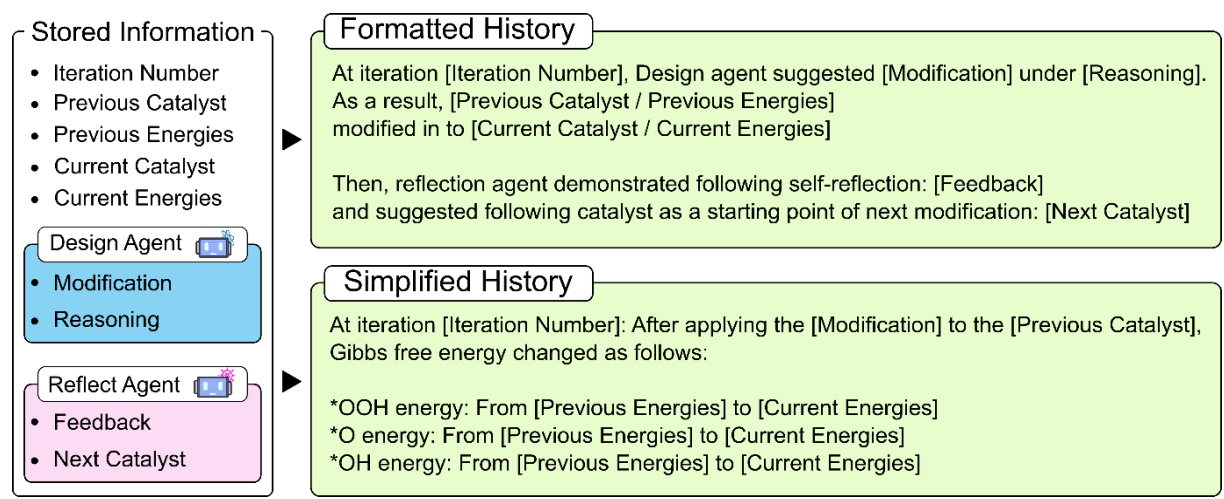

**Figure S2.** Schematic of the formatted and simplified history representation derived from the information stored at each design step.

**Supplementary Note B. Details of Pre-validation for MLFF and LLM**

**B.1 Dataset**

In this study, we employed pre-trained UMA as MLFF surrogate for DFT. Because SAC system was not included in the training data of UMA, it was necessary to assess its optimization reliability for SACs prior to application within the design framework. Moreover, as no publicly available SAC benchmark dataset suitable for this purpose exists, we constructed a dedicate validation dataset by performing DFT calculations on structure representative of those expected to emerge during the catalyst design loop.

For this validation dataset, geometry optimizations were carried out for M-$N_4$ SACs with nine different transitions metals (Co, Cu, Fe, Ir, Mn, Ni, Pd, Pt, Ru) as the center atom, considering three intermediates (*OOH, *O, *OH) as well as configurations in which an additional axial ligand (OH) binds simultaneously with the adsorbate. This resulted in a dataset comprising 3,107 optimization images, corresponding 3,107 DFT total energies, 579,057 atomic forces and 54 binding energies. To further evaluate prediction accuracy under variations in the local environment of the binding site, additional datasets were generated for $CoN_4$ systems featuring COC or COH functional groups on the second shell of carbon support, as well as structures in which first shell N atom were substituted with hydrogen or removed to form defects. This data construction yielded an additional 420 DFT total energies and 76,098 atomic forces.

**B.2 Pre-validation of MLLFF**

Validation was primarily performed using the UMA with the 'OC20' domain, which is weighted toward heterogenous alloy catalysis systems and was also employed in the actual design framework. For comparison, validation was additionally conducted using UMA with the 'OMat' domain, which is biased toward crystalline bulk materials. The results revealed that UMA with the 'OMat' domain exhibited substantially inferior prediction performance (**Figure S3**) compared to the model weighted on 'OC20'. In contrast, the UMA with 'OC20' maintained high auccracy even for SACs with modified local environments (**Figure S4**)

Notable, the validation datasets include systems that were absent from the UMA training data, such as two-dimensional materials with functional groups or defects, octahedral geometrics involving simultaneous ligand and adsorbate binding and *OOH adsorbate. Considering these out-of-distribution characteristics, the performance demonstrated by UMA in this study is particularly significant, supporting its suitability as a surrogate model.

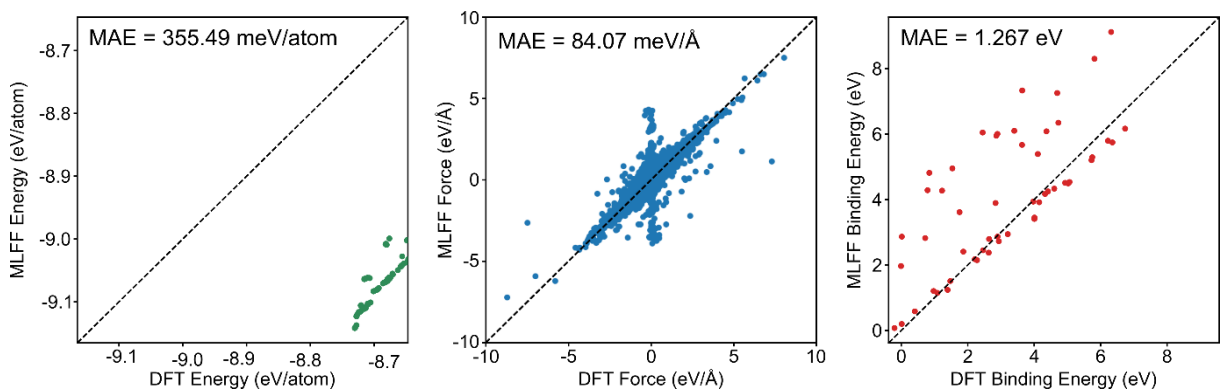

**Figure S3.** Parity plots between MLFF-predicted and DFT-calculated energies per atom, atomic forces and binding energies of pristine M-N$_4$ SAC system. In this prediction, UMA with OMAT task is used as MLFF.

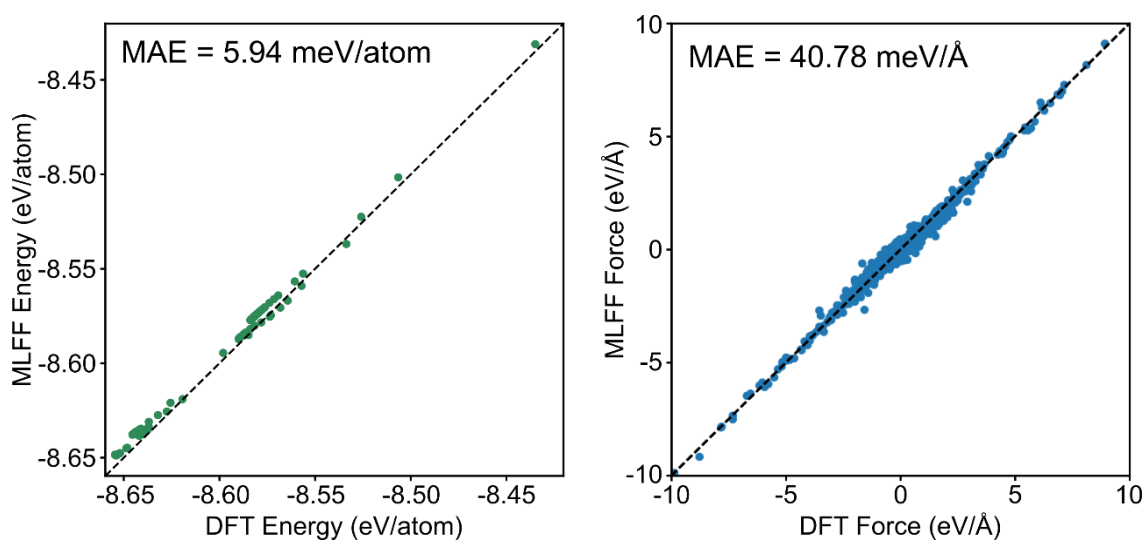

**Figure S4.** Parity plots between MLFF-predicted and DFT-calculated energies per atom and atomic forces of modified M-N$_4$ SAC system. In this prediction, UMA with OC20 task is used as MLFF.

### B.3 Pre-validation of LLM

In most cases, the modifications proposed by the LLM induced changes in binding energies consistent with both the intended direction and the underlying reasoning provided by the model. Nevertheless, a limited number of discrepancies were observed. The first type of inconsistency arises from structural distortions driven by local environment effects, as illustrated in **Figure S5a**. In this case, a functional group located in the second shell migrated toward the metal center during MLFF-optimization. This migration destabilized the bare catalyst structure, increasing its DFT total energy and consequently leading to a reduction in the binding energy. Although the LLM correctly anticipated the change in electron density, it failed to predict binding energy trend due to the unforeseen structural rearrangement.

A second source of inconsistency originates from incorrect analogies to prior cases, as shown in **Figure S5b**. Even when identical modifications are applied, the resulting binding energy trends can differ depending on the current condition of catalyst. In such instances, the LLM neglected this contextual dependence, leading to erroneous predictions for both binding energy and electronic density changes.

To mitigate these limitations within the proposed framework, we introduced additional mechanisms to inform the LLM when substantial structural rearrangements occur after geometry optimization, such as changes in binding configurations. Furthermore, modifications were recorded in the design history together with updated textural descriptions of the catalyst structures. This approach enables LLM to associate binding energy trends with both the catalyst environment and the applied modification, thereby improving its contextual reasoning in subsequent design iterations.

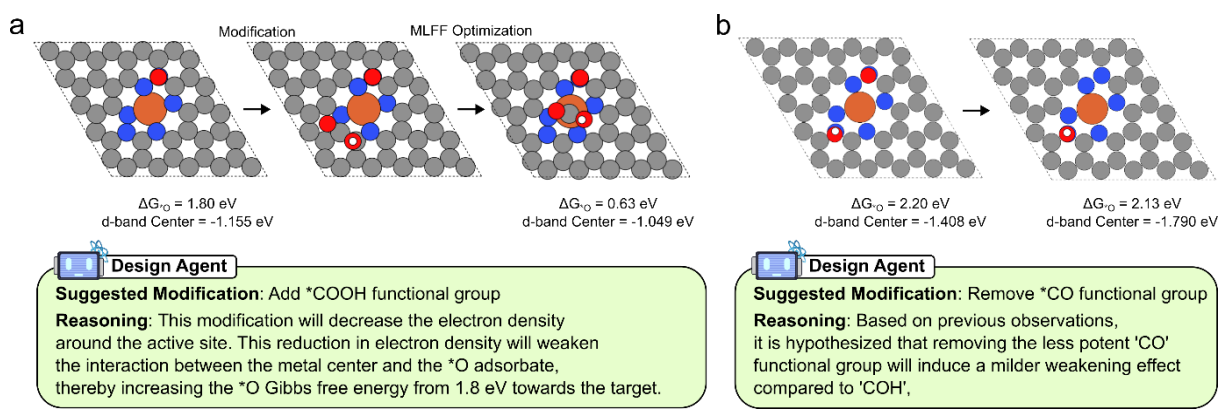

**Figure S5.** Examples of discrepancies between reasoning of the design agent and calculated results. (a) Case of binding site change due to movement of functional group. (b) Case of incorrect reference to previous observation.

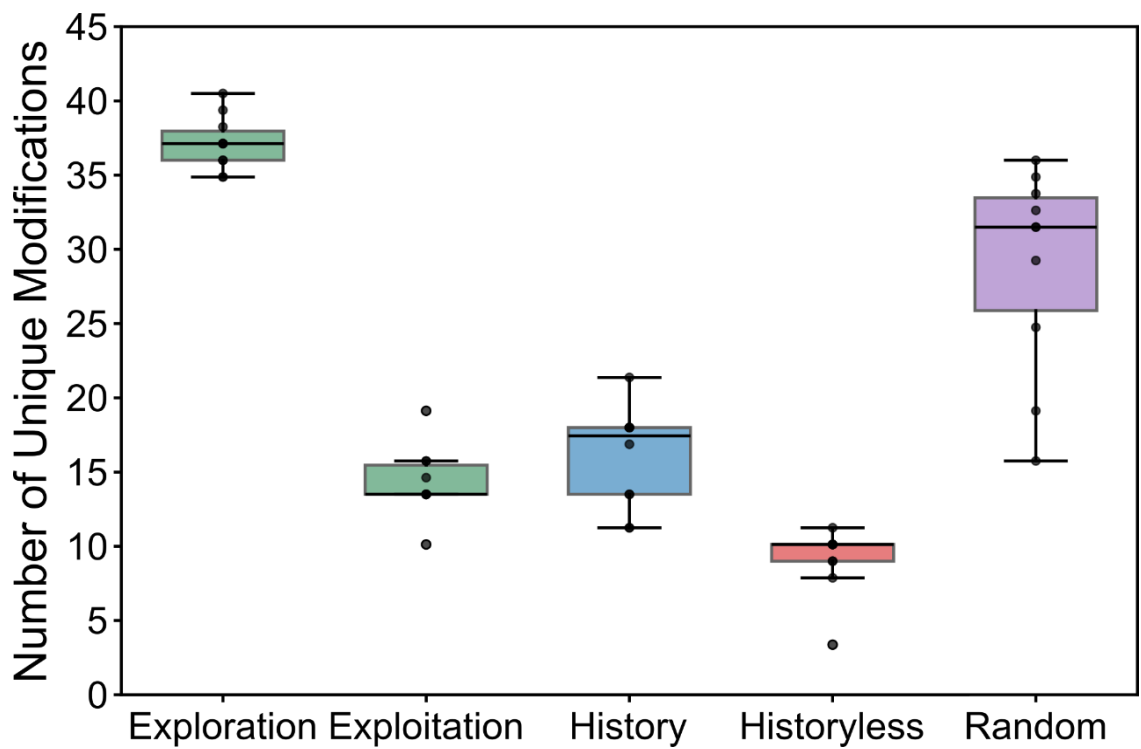

**Figure S6.** An example of exploration summary report from exploration report agent.

**Figure S7.** Number of the unique modifications proposed during the 50 design steps, averaged over 10 independent design runs. The labels "Exploration" and "Exploitation" correspond to the exploration and exploitation phase of *history + exploration* strategy, respectively,

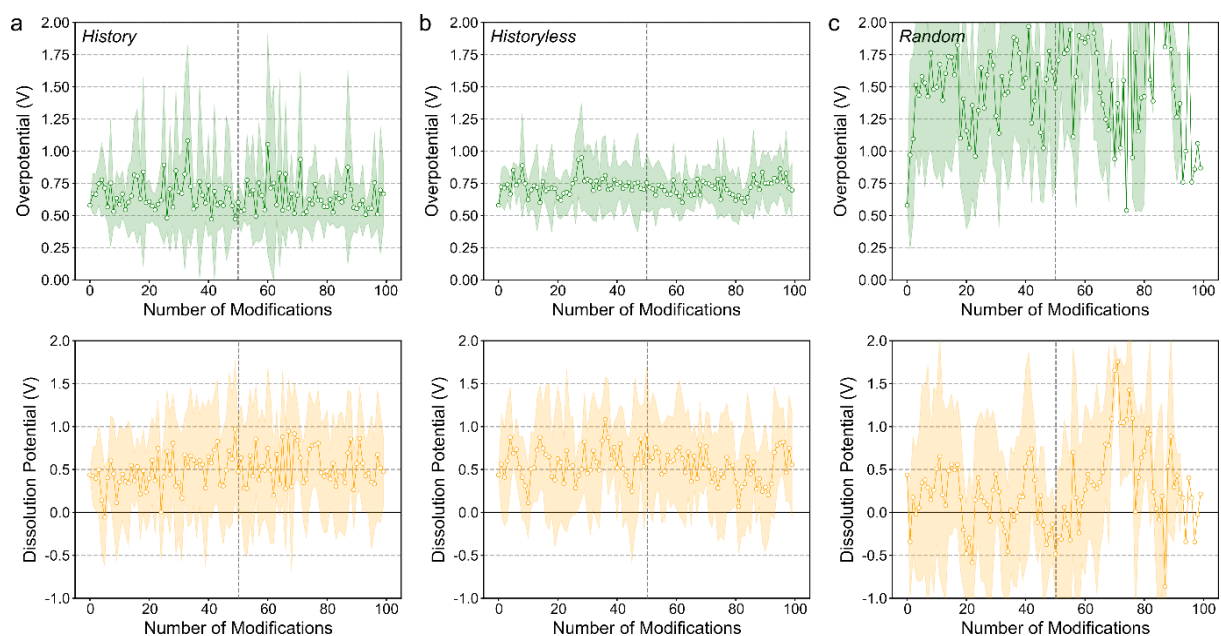

**Figure S8.** Evolution of overpotential and dissolution potential averaged over 10 independent design runs using (a) *history*, (b) *historyless* and (c) *random* strategy.

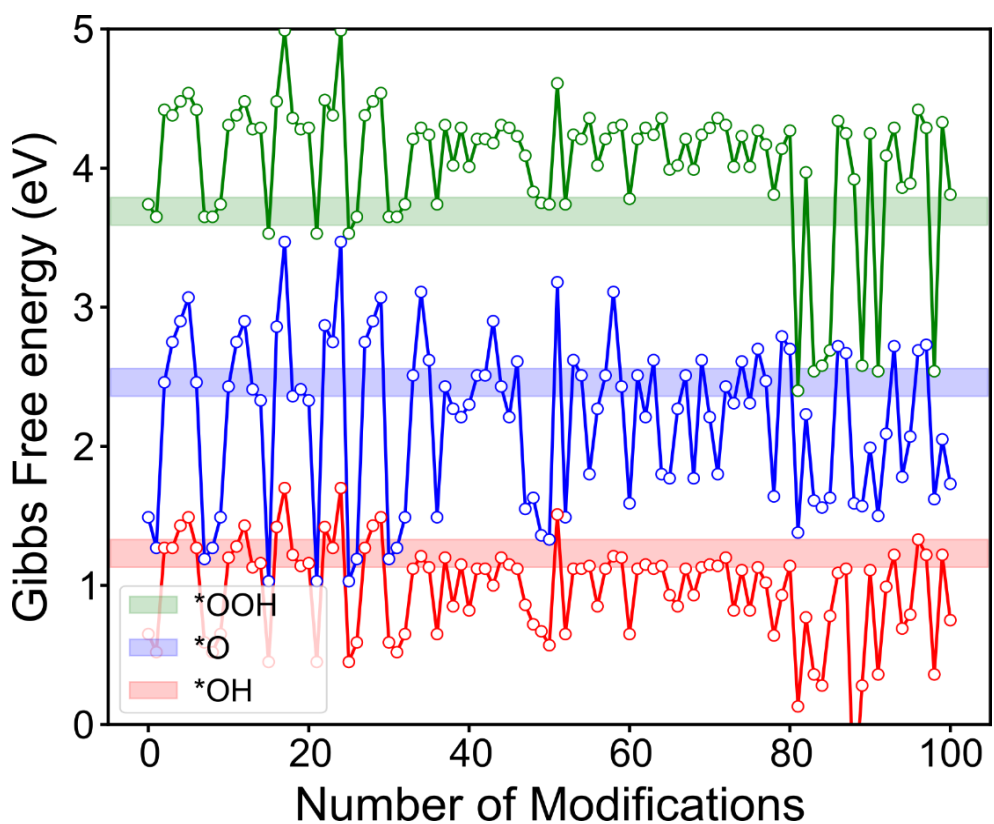

**Figure S9.** Evolution of the Gibbs free binding energies of *OOH, *O and *OH over 100 design steps from a design run that failed to break the scaling relations. Shaded regions denote the optimal energy windows required to achieve and overpotential below 0.1 V. Because the binding energies increase and decrease concurrently according to the scaling relations, the three intermediates cannot simultaneously fall within their respective optimal ranges.

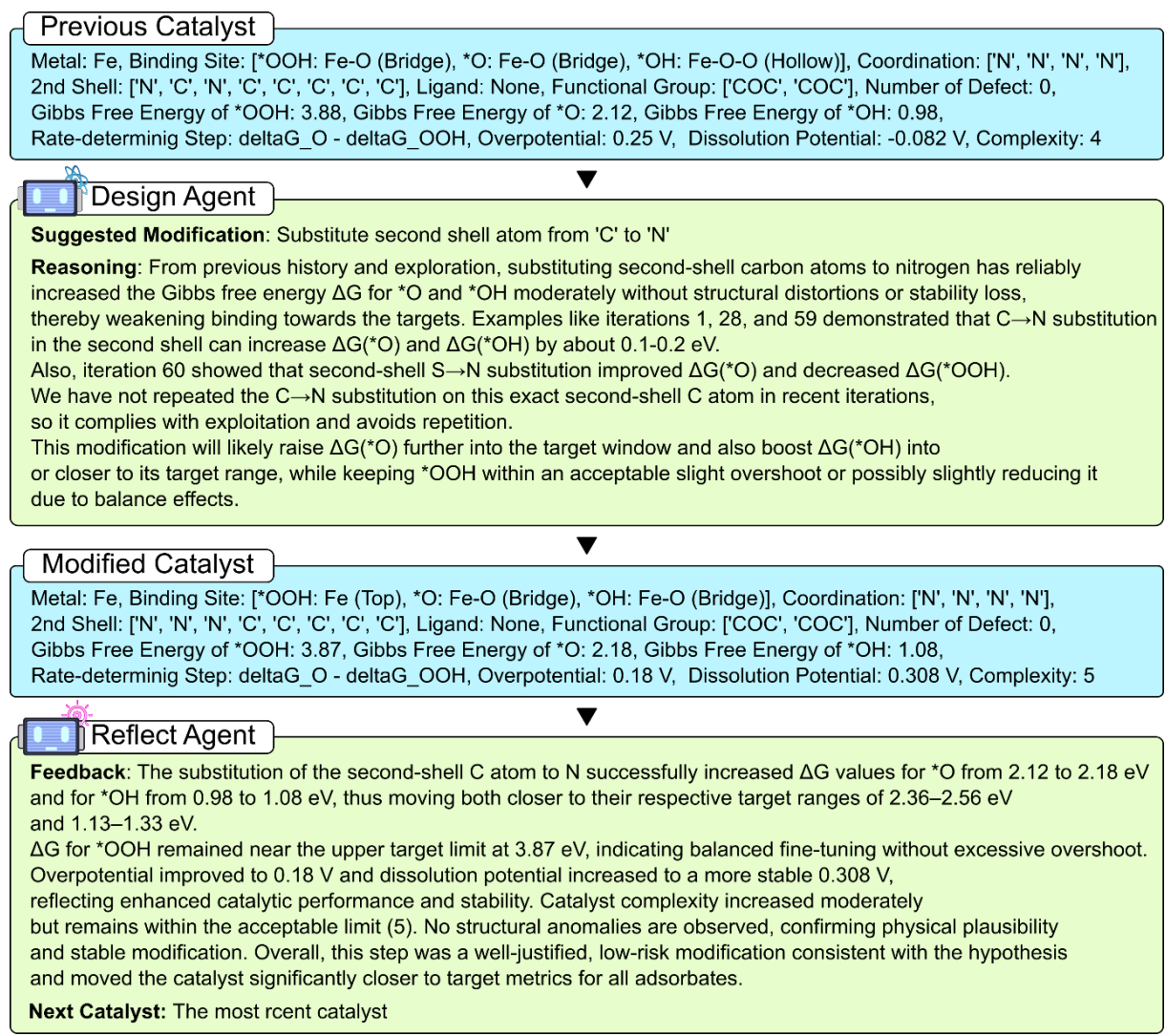

**Figure S10.** An example of design and self-reflection workflow leading to a minimum overpotential (0.18 V). The design agent suggests a modification by leveraging accumulated design history to selectively tune binding energies. The reflect agent subsequently evaluates modification and calculation results, confirms its effectiveness and provide feedback to guide the next design step.

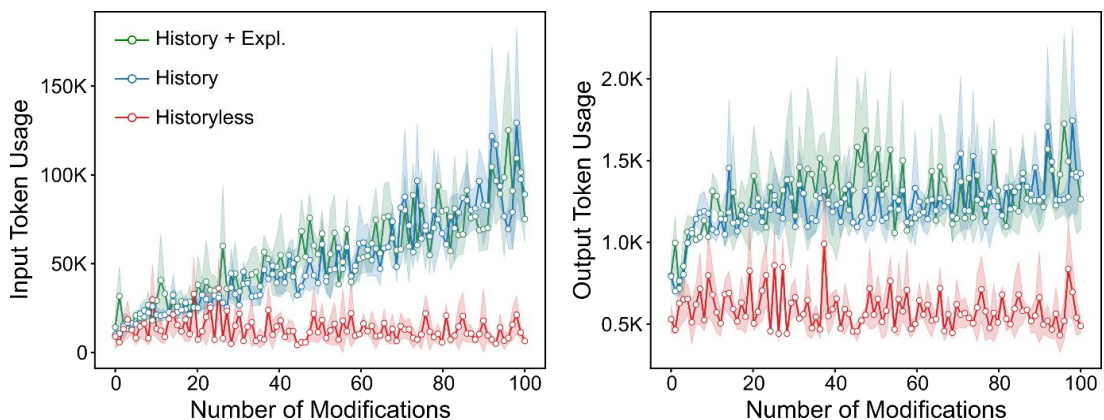

**Figure S11.** Input and output token usage changes by progress of design run. The *historyless* strategy maintains nearly constant input token usage comparable to the initial value, whereas the *history* and *history + exploration* strategies exhibit a gradual increase in input token usage as design history accumulates.

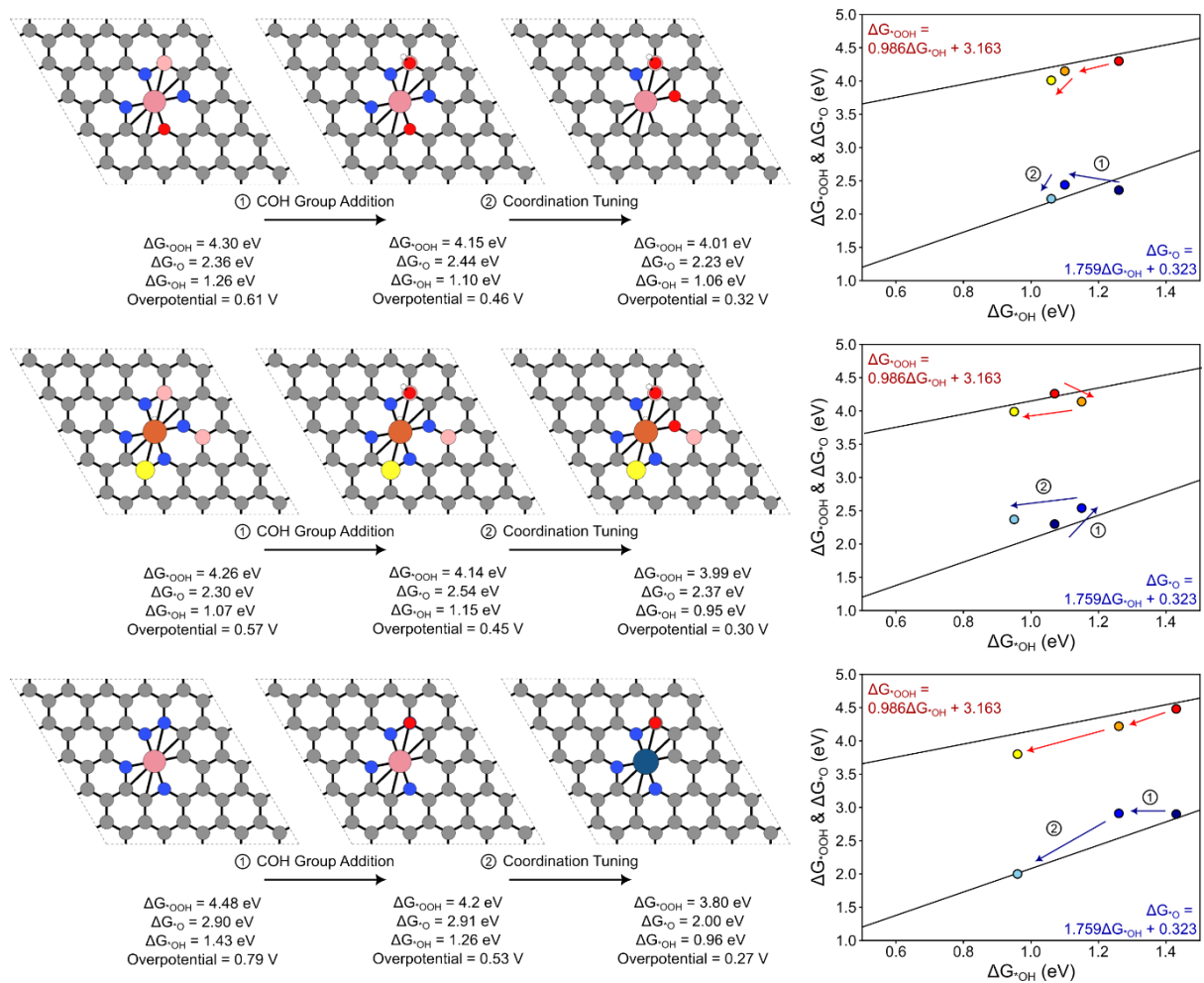

**Figure S12.** Various representative modification process to design promising catalysts by breaking scaling relations and lower limit of overpotential.

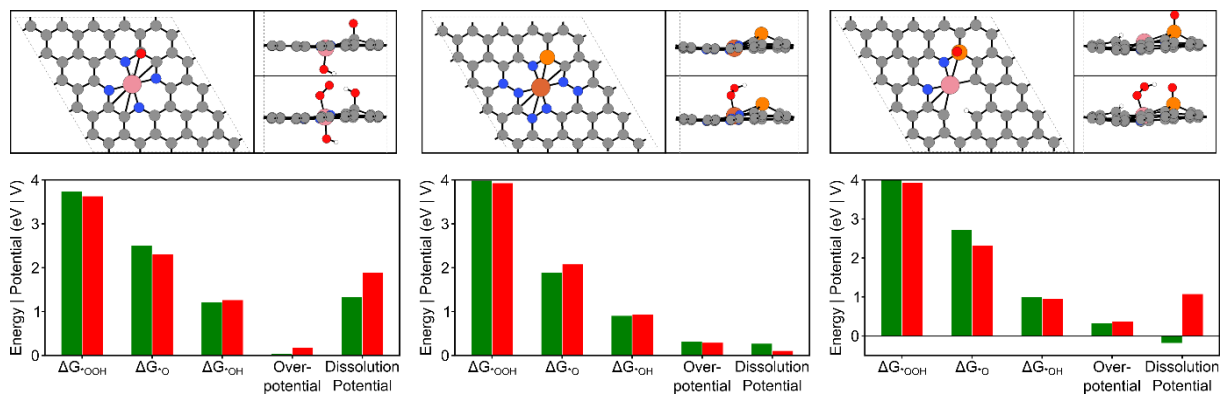

**Figure S13.** Representation of one of high-performance catalyst structures discovered during the design runs, together with a comparison of MLFF-predicted and DFT-calculated performance metrics. Data includes Gibbs free binding energies, overpotential ($\eta$) and dissolution potentials ($U_{diss}$) for each candidate.

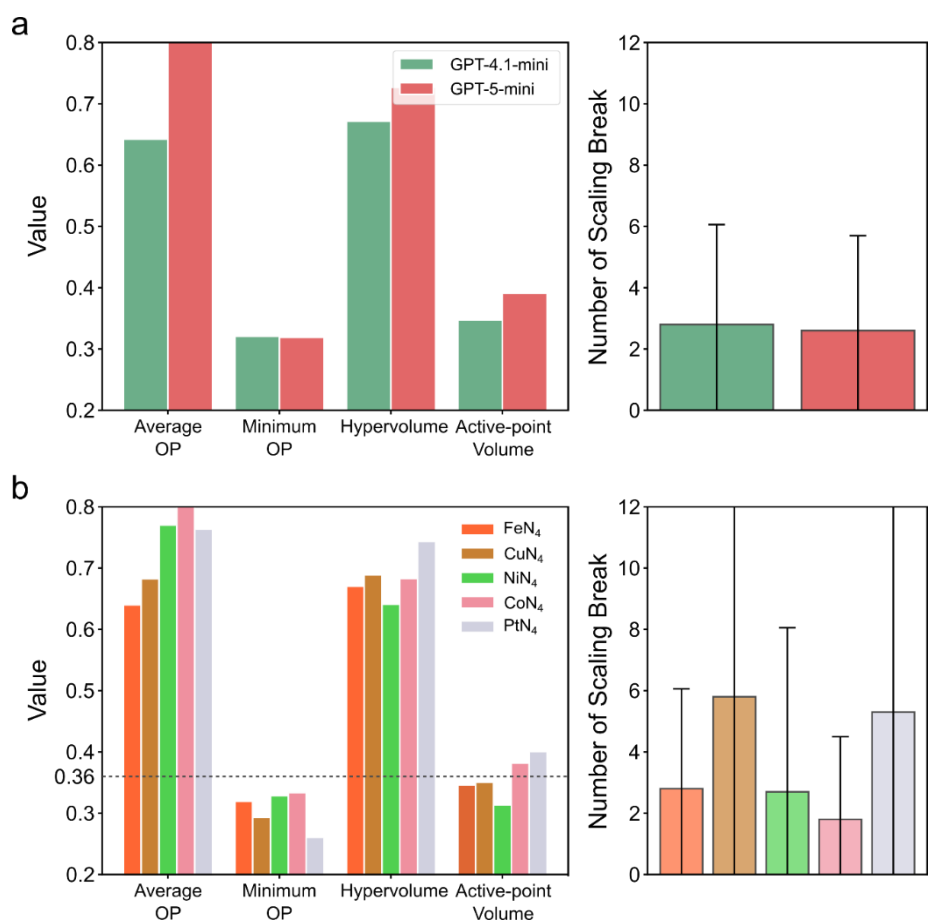

**Figure S14.** Comparison of overall design performance among (a) different LLMs and (b) different starting SACs. The dashed line indicates lower limit of overpotential defined by the scaling relations.

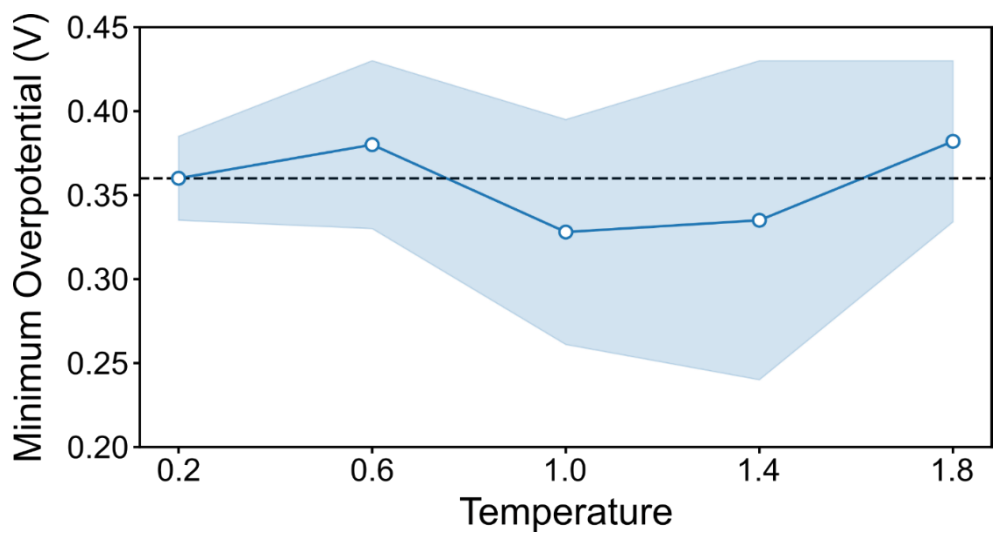

**Figure S15.** Minimum overpotentials averaged over 10 design runs with each temperature setting. The dashed line indicates lower limit of overpotential defined by the scaling relations.

**Table S8.** The Gibbs free energy correction values for adsorbates and gaseous molecules. All units are in eV.

| Adsorbates | ZPE | ∫C$_p$dT | -TS | Molecules | ZPE | ∫C$_p$dT | -TS |
|---|---|---|---|---|---|---|---|
| O* | 0.08 | 0.03 | -0.06 | H$_2$ | 0.28 | 0.09 | -0.41 |
| OH* | 0.37 | 0.04 | -0.07 | H$_2$O | 0.57 | 0.10 | -0.67 |
| OOH* | 0.44 | 0.05 | -0.09 | | | | |